\documentclass[%
amsmath,amssymb,
]{revtex4-2}

\usepackage{graphicx}% Include figure files
\usepackage{dcolumn}% Align table columns on decimal point
\usepackage{bm}% bold math
\begin{document}

\preprint{APS/123-QED}

\title{Feasible Architecture for Quantum Fully Convolutional Networks}% Force line breaks with \\
%\thanks{A footnote to the article title}%

\author{Yusui Chen}
\email{yusui.chen@nyit.edu}
\affiliation{%
Physics Department, New York Institute of Technology, Old Westbury, NY 11568, USA
 %This line break forced with \textbackslash\textbackslash
}%
%\altaffiliation[Also at ]{Physics Department, XYZ University.}%Lines break automatically or can be forced with \\
\author{Wenhao Hu}%
% \email{Second.Author@institution.edu}
\affiliation{%
School of Computing Science, University of Glasgow, Glasgow G12 8QQ, UK
}%

%\collaboration{MUSO Collaboration}%\noaffiliation

\author{Xiang Li}
% \homepage{http://www.Second.institution.edu/~Charlie.Author}
\affiliation{
 QuantumX Technologies Inc., 100 Wall Street No. 1602, New York, NY 10005 % with \\
}%

\date{\today}% It is always \today, today,
             %  but any date may be explicitly specified

\begin{abstract}
Fully convolutional networks are robust in performing semantic segmentation, with many applications from signal processing to computer vision. From the fundamental principles of variational quantum algorithms, we propose a feasible pure quantum architecture that can be operated on noisy intermediate-scale quantum devices. In this work, a parameterized quantum circuit consisting of three layers, convolutional, pooling, and upsampling, is characterized by generative one-qubit and two-qubit gates and driven by a classical optimizer. This architecture supplies a solution for realizing the dynamical programming on a one-way quantum computer and maximally taking advantage of quantum computing throughout the calculation. Moreover, our algorithm works on many physical platforms, and particularly the upsampling layer can use either conventional qubits or multiple-level systems. Through numerical simulations, our study represents the successful training of a pure quantum fully convolutional network and discusses advantages by comparing it with the hybrid solution.
\end{abstract}

%Use showkeys class option if keyword
                              %display desiredhttps://www.overleaf.com/project/60ad742a5a37694d8838a30d
\maketitle
 
%\tableofcontents
\section{Introduction}
% ML -> CNN
Deep learning as a method of data analysis allows computers to discover
%? 
and improve 
the model from data and perform automatically with minimal human intervention. Convolutional neural networks (CNN) \cite{CNN1,CNN2} as one type of deep learning algorithms have many successful applications in the field of science and technology, e.g., high-energy particle physics, condensed matter physics, biological and chemical systems, image recognition, natural language processing\cite{cnn3,cnn4,cnn5,cnn6,cnn7}. Through the multiple designed convolutional and pooling layers, the original data is coarse-grained and fully connected. As a result, CNNs provide a practical method to capture the spatial and temporal dependencies. 

% CNN -> QCNN 
However the success of deep learning algorithms highly depends on the volume of data and the computational power. Quantum machine learning (QML) is a feasible solution to address the challenge of solving large data involved computational problems because quantum computing has its natural advantage over the classical ones that it can turn the dense classical computation into a series of measurements on the quantum system and speedup the process exponentially \cite{nielsen_chuang_2010,seth2011,seth2017,qa-fiance,qa-covid,Qa-photon,qa-Carolan2020,qa-atoms,qa-lattice,qa-chemistry,qa-mcclean2014,rupack}. Due to the lack of quantum error correction\cite{chen2014,chen2014_1}, current quantum computers cannot implement the generic quantum algorithms, e.g., Shor algorithm, Grover algorithm, on the noisy intermediate-scale quantum (NISQ) devices\cite{preskill2018}. But recent works have demonstrated that the variational quantum algorithms (VQAs), a method using the variational principle to provide approximating solutions to a computational problem, can be implemented on NISQ devices. In general, a VQA can be mapped to a fully-parameterized quantum circuit (PQC) and driven by a classical optimizer\cite{QVA-NISQ,QVA-cerezo,QVA-highEnergy,QVA-ibm,QVA-McClean,QVA-tiangong,QVA-Zhu2019}. Quantum convolutional neural networks (QCNN) as an example of QML have emerged as the overlap of classical convolutional networks (CNN) and QML, and provide a potential solution to speeding up the data processing an·d increasing the capability of handling data base on the NISQ devices\cite{cong,QCNN1,qcnn2,QCNN3}. In addition, QCNNs provide us the potential of further quantum supremacy applications in deep learning area, because CNNs are the base architecture of many advanced neural networks. 
% QCNN ->QFCN
Although the architecture of CNN/ QCNN provides the coarse-gained model and successes in pattern recognition, it has the limited ability to perform the dense multiple data patterns recognition. Fully convolutional networks (FCN) \cite{FCN0,fcn1,fcn2,fcn3,FCN4} as a natural extension from the CNN architecture to the encoding-decoding framework can realize the complex data pattern discovery and supply a practical method to perform the dense prediction that label each unit of data with a specific class, e.g., semantic segmentation, image segmentation, carrier signal detection in broadband power spectrum, and time series classification. A compromise solution to perform FCNs in the framework of quantum computing is a hybrid model in which the outcomes of QCNN are fed into a classical FCN. But the hybrid solution cannot avoid bottlenecks in classical algorithms. As a result, a pure quantum model is necessary that a PQC consists of a pure quantum CNN and the quantum upsampling (QFCN). Moreover, the quantum solution can be performed on the one-way quantum computer which allows dynamical programming and reuses the measured qubits to perform the decoding process. Particularly, in the domain of computer vision, video recognition with deep learning is constructed to handle continuous data with time. 

Inspired by QCNNs and the multi-scale entanglement renormalization ansatz (MERA)\cite{phy-vidal,phy-vidal2}, we present a fully PQC for the QFCN model to perform semantic segmentation on classical data. Although VQA-based algorithms have been proposed to address the classical hard computational problems on quantum computers, some key features such as trainability and efficiency of VQAs are still under debate. In this work, we also compare the performance of the hybrid model and the pure quantum one, and discuss the impacts of volume of data set and the fine adjustments of the decoding algorithm. 

In this work, we start by introducing the basic PQC for QFCN model based on the architecture of VQAs. We then focus on the numerical simulations of the algorithm in different setups, including the hybrid style and the pure quantum structure, as well as the numerical stability depending on different data sets. In the last section, we present predictions for quantum advantage on NISQ devices and conclude on the possible applications in other relevant quantum computing fields. 

\section{Theory}
\subsection{Quantum variational learning}

\begin{figure}[htp]
\includegraphics[width=3.2 in]{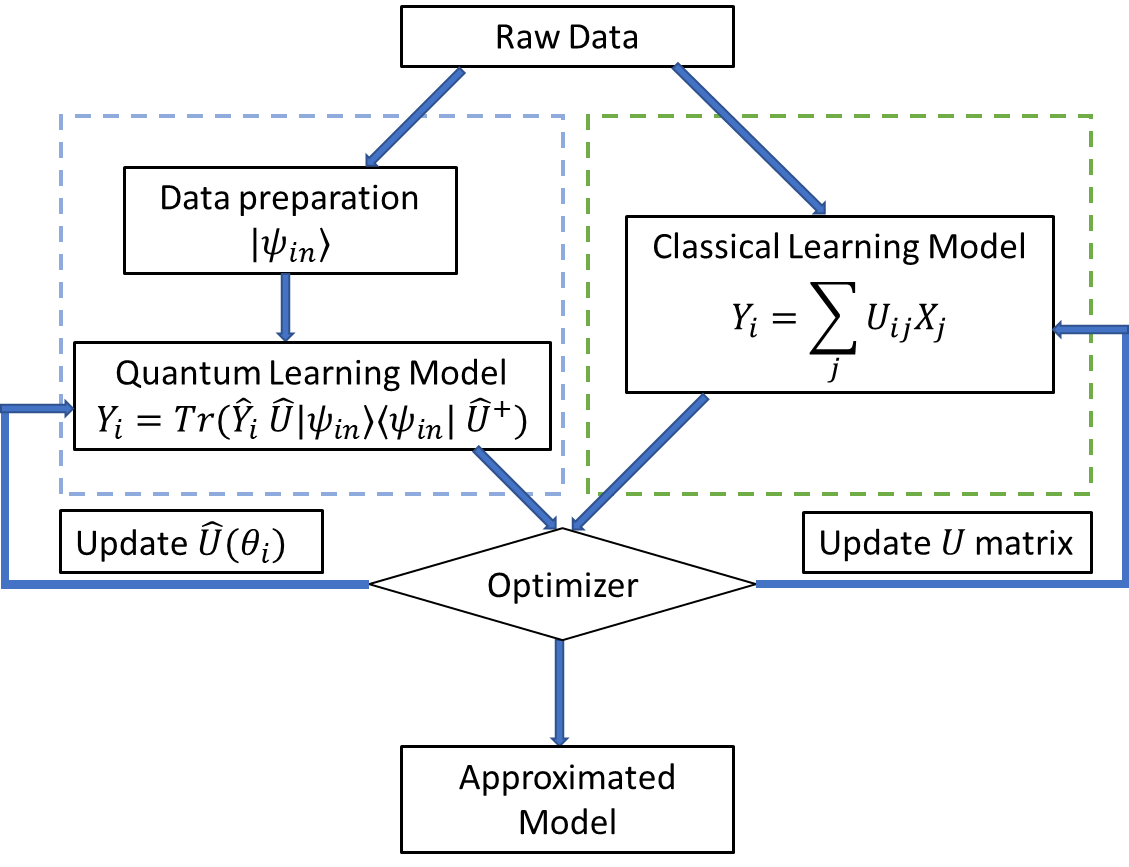}
\caption{\label{fig:general} Scheme of generic quantum and classical variational learning algorithms. }
\end{figure}

Generally, the variational learning is training a fully-parameterized map $f:X\mapsto Y$ that minimize the \textit{distance} between the trained model and the true one. Starting from a set of random chosen parameters, the optimizer evaluates the quality of the model by measuring the distance between the output and the labeled data in the test group and update a new set of parameters in the model. Once the distance is less than the error tolerance, such a model is considered as the approximated-real model. In current techniques, 
% change to "the bottlenecks are that"
the most time-consuming components are that: (1) calculating the output variable due to the large size of the data; (2) finding the global minimal position on the high dimensional parameter hyper surface is difficult\cite{benenti,qa-mcclean2014}.

 A typical quantum variational algorithm (QVA) is consisting of three steps: (1) preparing the raw data in the initial state $|\psi_{in}(\vec{x})\rangle $; (2) measuring the output state $|\psi_{out}\rangle= \hat{U}(\theta )|\psi_{in}\rangle$ to \textit{compute} the required data $Y_i = Tr(\hat{Y}_i|\psi_{out}\rangle \langle \psi_{out}|) $, where the operator  $\hat{U}(\theta)$ characterizes the whole quantum circuit and $\{\theta_i\}$ in $\theta$ are all possible parameters; (3) updating the parameters based on the outcome of the classical optimizer $\hat{U}(\theta) \rightarrow \hat{U}(\theta')$. As shown in Fig. \ref{fig:general}, QVAs can speedup the computing process by  turning the classical computing process into a series of quantum measurements. As a result, this advantage will be dominant when the size of data exceeds some thresholds. Some practical applications of quantum advantage over classical supercomputers have been explored, e.g., variational algorithms, machine learning problems.  
 
 In the classical optimizer, the \textit{distance} between the trained data $Y_i$ and the labeled data $\mathcal{Y}_i$ is characterized by the cost function. In this work, we use the mean of squared error (MSE)\cite{MSE} to serve as the loss function $\mathcal{L}(\theta)$,
 \begin{align}
     \mathcal{L}(\theta ) =  \frac{1}{N} \sum_{j=1}^N (\mathcal{Y}_j - Y_j)^2.
 \end{align}
Meanwhile, the loss function minimum can be located by tweaking the parameters $\theta$ iteratively, 
\begin{align}
    \theta^{k+1} = \theta^k - \delta \nabla_\theta \mathcal{L}(\theta^k), \nonumber
\end{align}
where $\delta$ is the step size, until the error tolerance ($\epsilon$) converges
\begin{align}
    |\mathcal{L}(\theta^{k+1}) - \mathcal{L}(\theta^k)| \le \epsilon. \nonumber
\end{align}

\subsection{Basic quantum CNN}
\begin{figure}[ht!]
\includegraphics[width = 3 in]{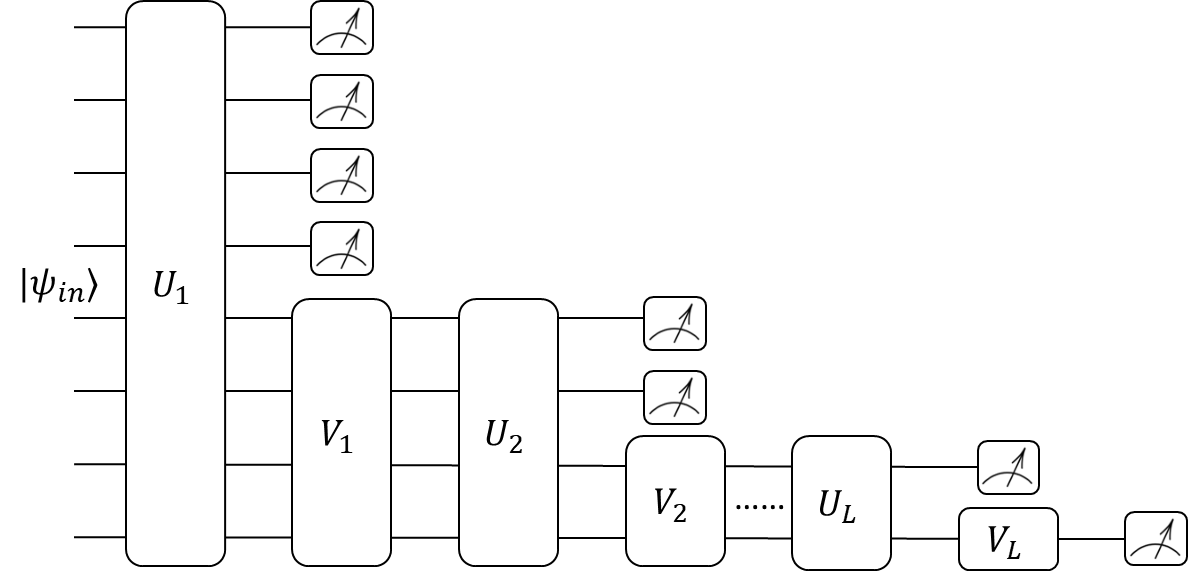}% Here is how to import EPS art
\caption{Sketch of multi-layer (L layers) quantum convolutional neural networks (QCNNs). All $\{U_i, V_j\}$ are linear combinations of two-qubit quantum gates which fully connect all engaged qubits. The size is progressively decreased by measuring some qubits. The network output is read out via measurements on the final qubit.}\label{fig:qcnn}
\end{figure}

Quantum convolutional neural networks (QCNNs) are originally aimed to classify the quantum phase of a given state in spin-chain models, which is a ubiquitous difficult question in many-body physics. In most recent proposals, QCNNs refer to a full-parameterized quantum algorithms which can be trained by a classical optimizer. In those models, as shown in Fig. \ref{fig:qcnn}, the neural network structure is a parameterized quantum circuit, which contains multiple quantum gates to build up a quantum computing mission. By naturally mapping the classical convolution computing onto a many-body Hamiltonian, QCNNs resolve the difficulty and enhance the efficiency when deal with a large set of data according to the increased system size. 

A generative QCNN consists of two types of layers: the convolutional layer $\hat{U}_j$ and the pooling $\hat{V}_j$ layer. Inside each layer, the operator $\hat{U}_i$ or $\hat{V}_j$ can be decomposed as a set of gates, connecting all engaged qubits where the quantum gates can be properly chosen depending on the physical realizations. Theoretically, every quantum gate operated on qubits is equivalent to the combination of two fundamental gates: single-qubit gates and two-qubit entangling gates. The major difference between the two types of layers is that the convolutional layer does not change the dimension of data, while the pooling layer decreases the size of data. In quantum computing, decreasing the dimension of data is naturally realized by performing partial measurement,
\begin{align}
    \rho_{R} &= Tr_M(\rho_{R\otimes M}) \nonumber \\
    &= \sum_{n,j,k,l,m} \langle \psi_M^n| \rho_{jklm} |\psi_R^j\rangle|\psi_M^k\rangle \langle \psi_R^l|\langle \psi_M^m|\psi_M^n\rangle \nonumber \\
    &= \sum_{j,l} (\sum_n \rho_{jnln} )|\psi_R^j\rangle \langle \psi_R^l|,
\end{align}
where $|\rho_{R\otimes M}\rangle$ is the total quantum state of the system before the measurement that can be explicitly written in the product basis of remaining qubits $\{|\psi_R^j\rangle\}$ and the to-be-measured qubits $\{|\psi_M^k\rangle\}$, associated with coefficients $\{\rho_{jklm}\}$. After performing partial trace, the element in the outcome state is $\rho_{jl} = \sum_n\rho_{jnln}$.
The final classification step is readout by the measurement on the last qubit. 

\subsection{Quantum fully convolutional networks}

% 需要把hyb 需要先测再重表示； pure 直接输入， 至少加一个小测量在Classical，而且他们都是 FCN upsampling （小方块里）不是 FCN
\begin{figure}[ht!]
\includegraphics[width = 3 in]{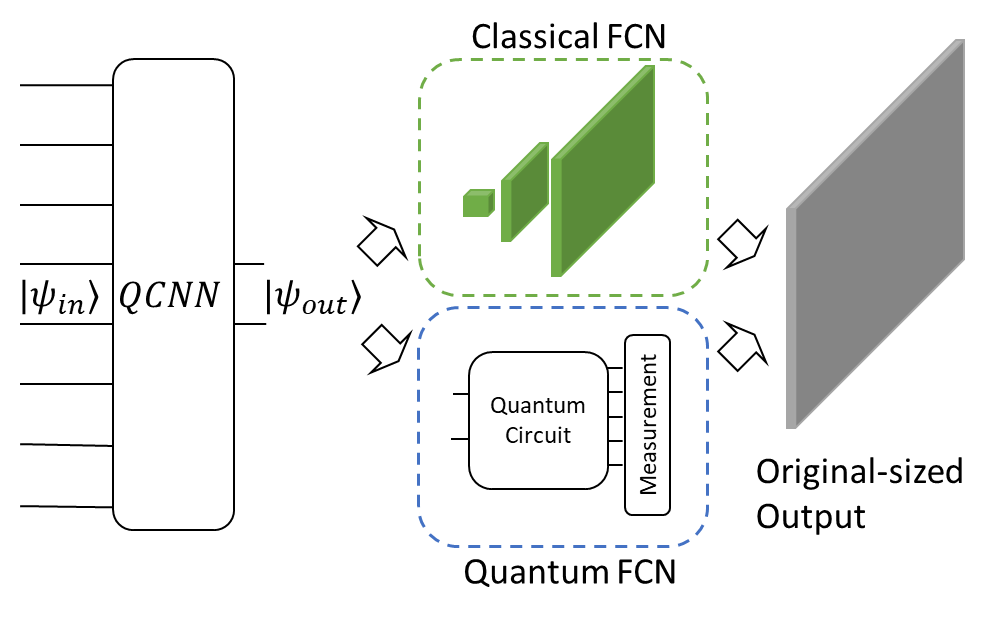}% Here is how to import EPS art
\caption{Sketch of two potential solutions to perform FCN based on the data obtained from the QCNN.}\label{fig:hybridpure}
\end{figure}

After extracting convoluted features and pooling, the coarse-grained data is fed into the FCNs to perform semantic segmentation. Different from the traditional CNN, the readout is used to \textit{learn} the semantics and location jointly, in which the key difficulty is resolving the inherent coupling between semantics and location. As a result, the next step is to upscale the coarse-grained data back to the original size. There are two potential solutions: (1) the hybrid solution and (2) the pure quantum solution, as shown in Fig.\ref{fig:hybridpure}. The advantages of the pure quantum solution is similar to that of QCNNs, comparing to the traditional CNNs, that the natural parallel computing in a quantum circuit can exponentially speedup the computing process and extend the ability to deal with large-sized data. Moreover, from the previous discussion, the pure quantum solution can bring into more global coherence in the learning process which is important and helpful to increase the quality of machine learning algorithms. 

\begin{figure}[ht!]
\includegraphics[width = 3 in]{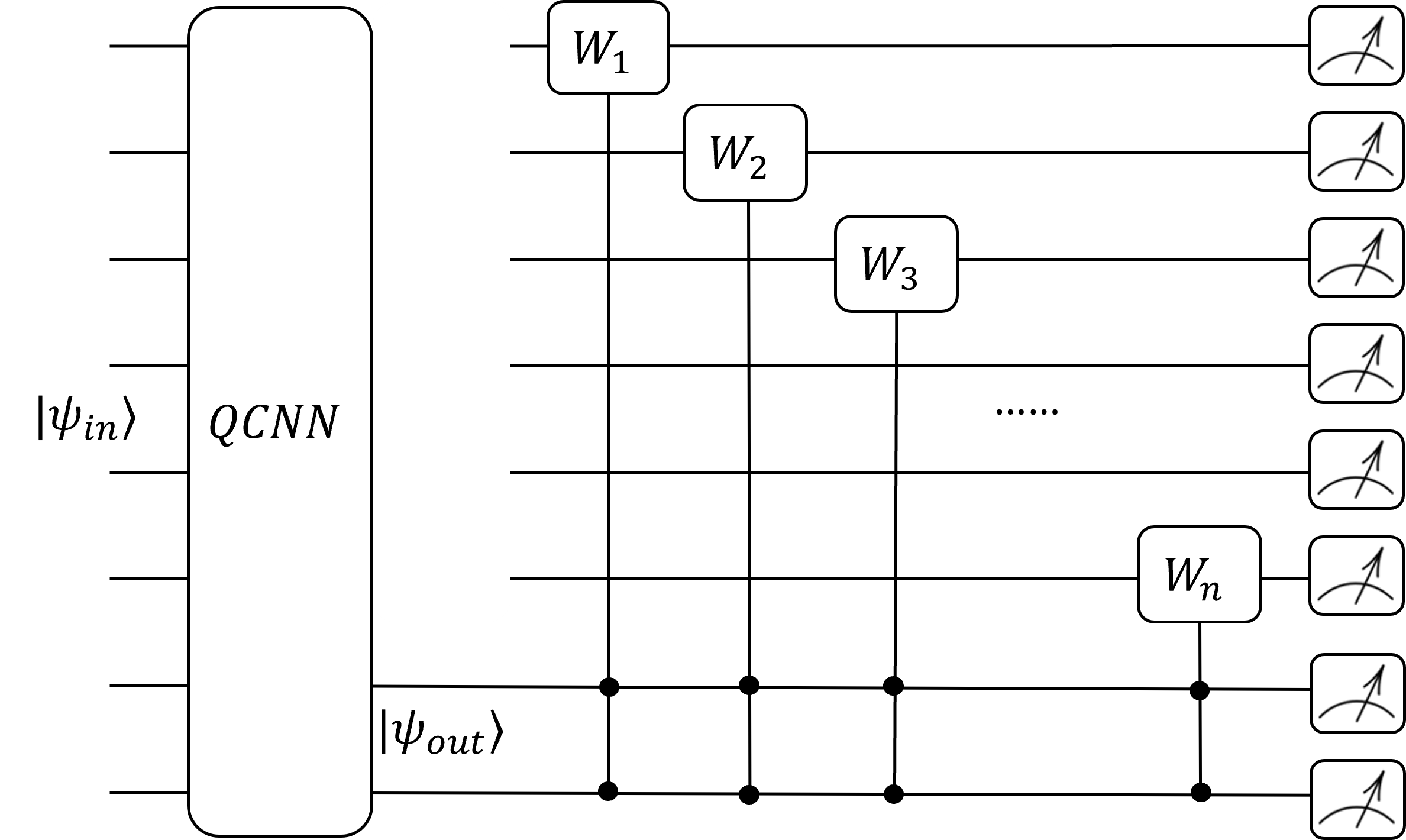}% Here is how to import EPS art
\caption{Sketch of quantum fully convolutional networks (QFCNs). All $\{W_i\}$ are linear combinations of two-qubit quantum gates which fully connect all engaged qubits. The size is increased by entangling measured qubits back into the circuit. The network output is read out via measuring all qubits.}\label{fig:qfcn}
\end{figure}

In this work, we propose a parameterized quantum circuit reproducing pure quantum fully convolutional networks (QFCNs), as shown in Fig.\ref{fig:qfcn}. In the classical FCN, the transposed convolution is realized by integrating The semantic vectors, the so-called upsampling layers, and extend  the dimension of data to the original size of the input data. As a result, the outcome can skip some intermediate steps in the deep learning and perform dense predictions. 

In the context of quantum computing, increase the dimension of states can be implemented by operating controlled gates on ancillary qubits. Such operations can extend the size of the Hilbert space of the entire system. Depending on the quantum system and the questions, the outcome from the QCNN is no longer restricted to the last readout qubit. As shown in Fig.\ref{fig:qfcn}, the readout can be multiple qubits and the upsampling layer can be realized 
\begin{align}
    |\psi_{out}\rangle \rightarrow
    \begin{bmatrix}
    I_{ctrl} & 0 \\
    0 & \hat{W}_j
    \end{bmatrix} \begin{bmatrix}
    |\psi_{out}\rangle \\
    |\psi_{an}\rangle
    \end{bmatrix}, 
\end{align}
where $|\psi_{an}\rangle$ is the prepared initial state of the ancillary qubits (we use the cluster state in our work). Following the original idea of VQAs, the upsampling layer $\hat{W}_j$ can also be composed using single-qubit gates and non-local two-qubit gates. Without loss of generality, we train the unique $\hat{W}_j$ for every ancillary qubit. In some cases, the model can be simplified by using limited number of $\hat{W}$ gates and lower the cost of computation. The to-be-determined parameters in $\hat{W}$ gates can be trained after multiple times of epochs.

Our QFCN architectures clone the classical FCNs in the quantum context, that re-distribute the global information stored in the outcome state of the QCNN circuit in the new system in the same size of the original data and reserve the inherent tension between global and local information via the entanglement. By measuring the outcome state, the learning process consists of initializing all parameters and progressively optimizing them
until convergence, 
\begin{align}
    \mathcal{L}(\theta) = \frac{1}{N}\sum_{j=1}^N |\vec{\mathcal{Y}}_j - \vec{Y}_j|^2,
\end{align}
where $\vec{Y}_j$ is the readout of all qubits in the end of upsampling layer and $\vec{\mathcal{Y}}_j$ is the column vector of the labeled data.

\section{Results}

\begin{figure}[ht!]
\includegraphics[width = 3 in]{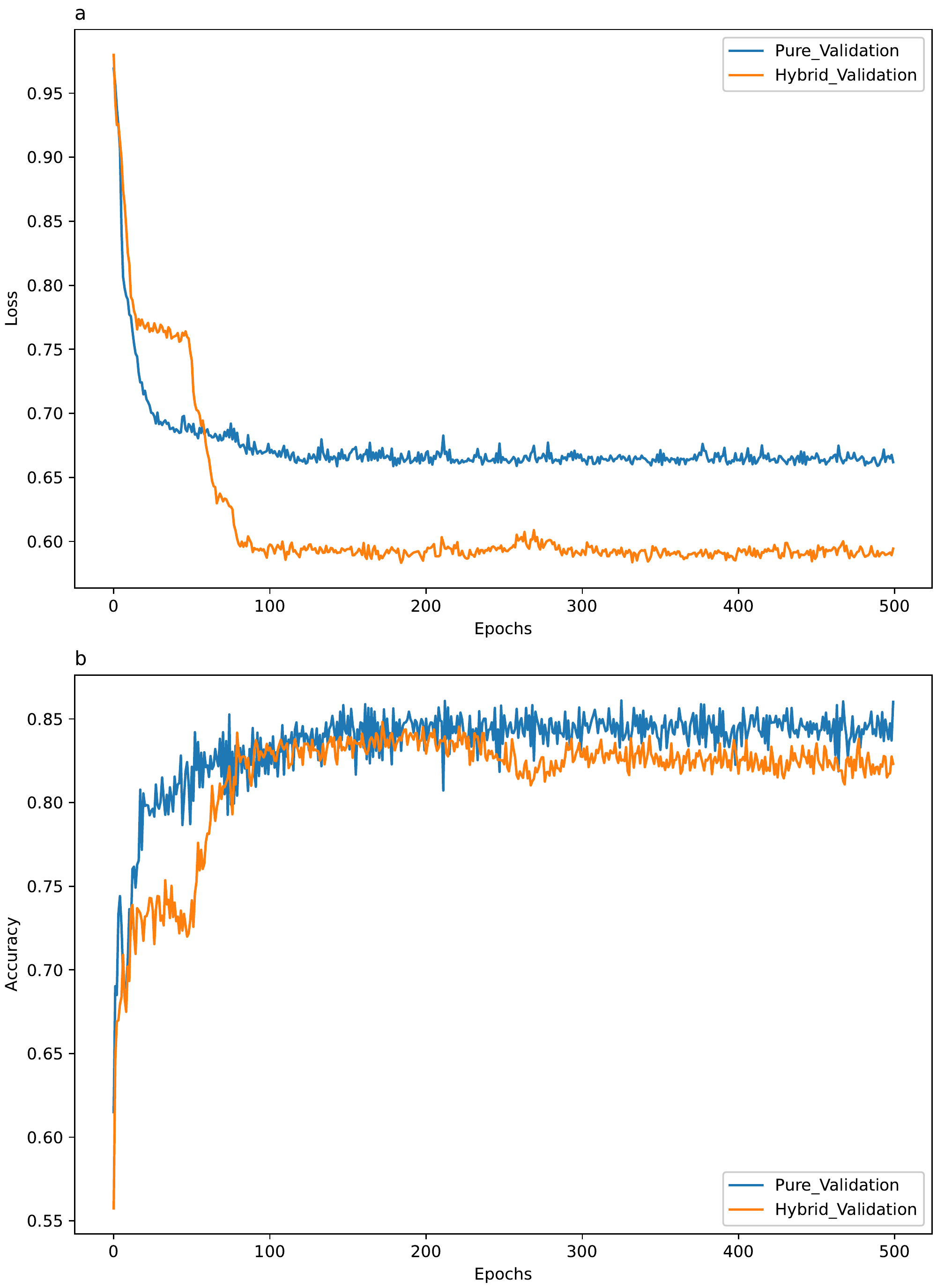}% Here is how to import EPS art
\caption{The performance of the pure quantum FCN (QFCN) model compared to the hybrid model.}\label{fig:hp}
\end{figure}

We simulate the quantum circuit consisting of 8 qubits using \textit{Google Tensorflow Quantum} package \cite{tensorflow}. The 8 qubits are initially prepared in the cluster state. In our simulations, each set of training data that carries noises and consists of two patterns is integrated in the 8 qubits as the rotation angle in the $RX(\theta)$ gate. In the convolutional and pooling layers, we employ 15 parameterized essential logic quantum gate to simulate every fully parameterized two-qubit gate \cite{cong}. For simplicity, all gates on every two qubits in the same layer are chosen to be the same. To perform upsampling, the output of the last pooling layer consists of two qubits. In Figs.\ref{fig:hp} (a) and (b), the results from the hybrid solution and the pure quantum solution are compared. It shows that the pure quantum solution converges faster than the hybrid solution. In addition, the accuracy of the trained model via pure quantum circuit is higher than that from the hybrid model. The comparison between the two potential solutions indicates that the quantum solution can better supply the global coherence between non-local qubits, which leads the higher accuracy in the validation. However, the trade-off is that the loss is not as low as the hybrid solution, as shown in Fig.\ref{fig:hp} (a), which explained that the classical FCNs can better expose the local patterns through fitting the training data. (Here, the overfitting issue is not considered.) The comparisons have validated that the overall QFCN architecture is performing as intended, that both the hybrid-and the pure- QFCN models can successfully realize the learning process respectively.

\begin{figure}[ht!]
\includegraphics[width = 3 in]{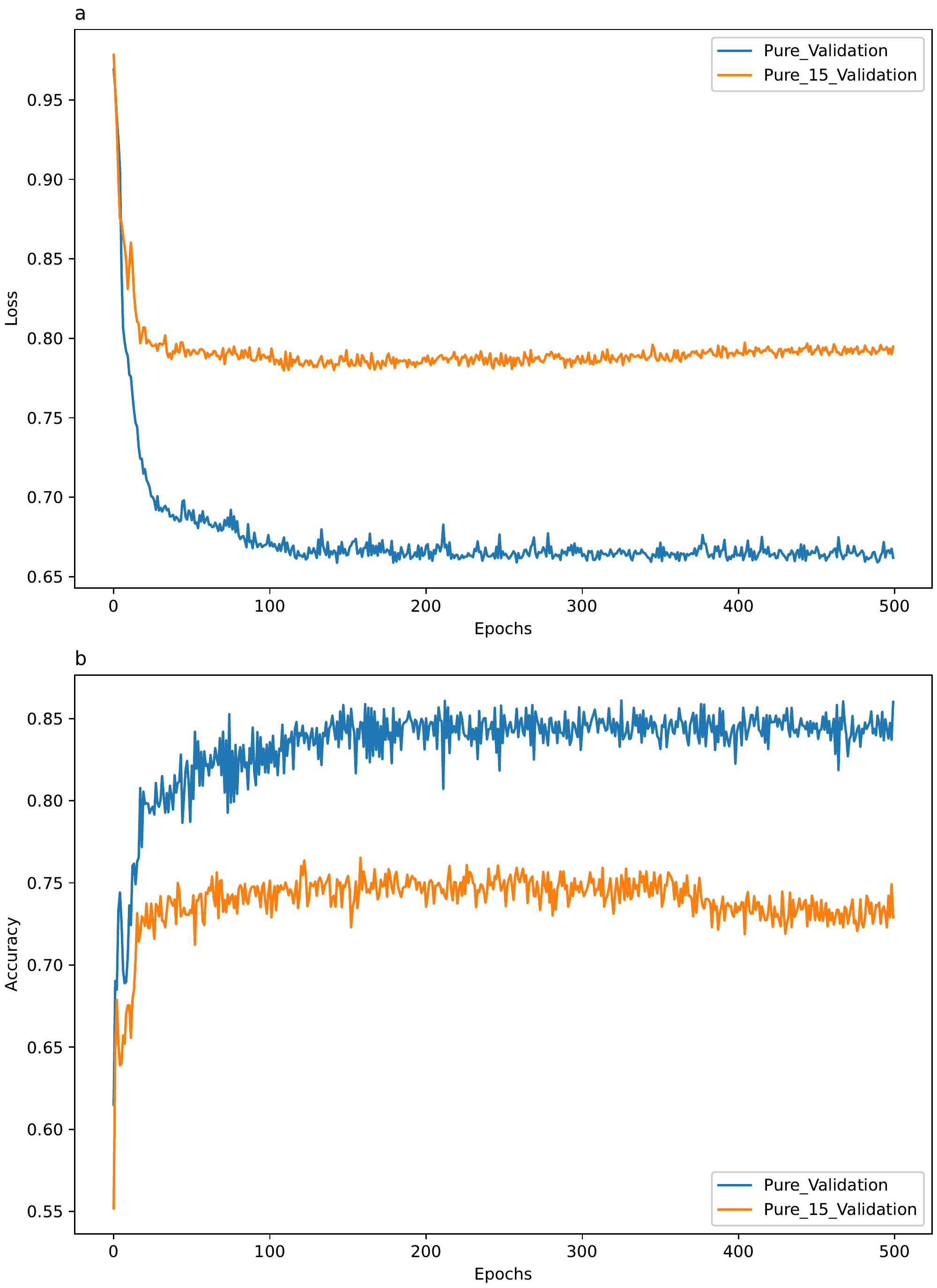}% Here is how to import EPS art
\caption{The performance of the unique upsampling gate model compared to the fully parameterized quantum gates model.}\label{fig:15}
\end{figure}

In our experiment, we also compare two possible ways to build up the pure quantum upsampling layers. Firstly, we operate a unique two-qubit gate to extend the size of the Hilbert space of the system, as what we set up in QCNNs. Secondly, we instead use different two-qubit gates for every two qubits pair. In Figs.~\ref{fig:15} (a) and \ref{fig:15} (b), the comparisons between the above-mentioned two setups are displayed. It is observed that in every way the fully-parameterized solution is better than the other that the fully-parameterized circuit converges faster and has better accuracy in validation and lower error in training model. In the classical algorithm, the upsampling kernels are initially same for every element in the data, which is similar to the operation that integrating an ancillary qubit into the system. The next step in the classical algorithm is to learn the variables in each kernel via gradient descent. It is worthy to note that each kernel is independent in this process. So it explains that the fixed and uniform ancillary qubit model does not work as it wipes out the independence of qubit at different locations. Moreover, in this experiment, using ancillary qubits to extend the size of the Hilbert space, but it is not the only solution. Other ancillary systems, e.g., multi-level systems, and continuous systems, can also serve as the upsampling kernel satisfying the deep learning process. It can extend our QFCN architecture workable on various quantum platforms for solving particular computational questions. 

\section{Conclusion}
In this work, we propose a quantum fully convolutional network to perform semantic segmentation on NISQ devices. Based on VQAs, the QFCN can be characterized by a parameterized quantum circuit that can be trained by converging the loss function following the classical optimizer. QFCNs can provide a promising, dynamical and scalable quantum machine learning application to speedup solving real-world problems. Our simulations prove the feasibility of performing classical FCNs on NISQ devices, and indicate that within a typical deep neural network architecture, QFCNs can increase accuracy in the network. Moreover, our results present some potential advantages of quantum algorithms: (1) pure quantum solutions can maximally speedup the computing process in the convolutional, pooling and upsampling layers; (2) quantum upsampling kernels can bring in the global coherence between non-local qubits which is better to fit large-sized data when there are weak couplings between separate parts. In addition, our algorithm offers a potential way to prepare measured qubits into a new initial state to perform QFCNs, a dynamical architecture that can increase the efficiency of the entire system. Lastly, our algorithm is open to arbitrary sized systems which serve as the upsampling kernel. As a result, we can freely choose the atomic system or continuous systems to realize upsampling, and use various quantum control strategies to operate the upsampling circuits. 

Although this research does not show the quantum advantage over the classical counterpart in the deep machine learning, the results indeed present that the quantum solution can achieve the convergence faster than the hybrid solution. And there are a number of practical questions need further discussions to make the QFCNs functionality, e.g., increasing the efficiency of preparing the initial states, performing measurements on qubits and operating the multiple two-qubit gates in the context of open quantum systems. These discussions will be included in our future works.

\begin{acknowledgments}
We acknowledge grant support from the NYIT's Institutional Support for Research and Creativity (ISRC) Grants.
\end{acknowledgments}

\bibliography{apssamp}

\end{document}